\documentclass[conference,a4paper]{IEEEtran}
\usepackage{amssymb}
\usepackage{amsmath}
\usepackage[dvips]{graphicx}
\usepackage{cite}
\usepackage{float}
\usepackage[T1]{fontenc}
\usepackage{array}
\usepackage{bm}
\usepackage{subfig}

\newtheorem{example}{Example}

\floatstyle{ruled}
\newfloat{algorithm}{tbp}{loa}
\floatname{algorithm}{Algorithm}

\begin{document}
\title{Joint Distributed Source-Channel Decoding for LDPC-Coded Binary Markov Sources}
\author{Reza~Asvadi\authorrefmark{1}, Tad~Matsumoto\authorrefmark{1}\authorrefmark{2}, and Markku~Juntti\authorrefmark{1} \\

\authorblockA{ \authorrefmark{1} Center for Wireless Communications (CWC), University of Oulu, Finland}
Emails: \{rasvadi, tadashi.matsumoto, markku.juntti\}@ee.oulu.fi
\authorblockA{ \authorrefmark{2} Japan Advanced Institute of Science and Technology, 1-1 Asahi-Dai, Nomi, Ishikawa,
Japan} Email: matumoto@jaist.ac.jp}

\maketitle
\renewcommand{\QED}{\QEDopen}

\begin{abstract}
We propose a novel joint decoding technique for distributed
source-channel (DSC) coded systems for transmission of correlated
binary Markov sources over additive white Gaussian noise (AWGN)
channels. In the proposed scheme, relatively short-length,
low-density parity-check (LDPC) codes are independently used to encode the bit
sequences of each source. To reconstruct the original
bit sequence, a joint source-channel decoding (JSCD) technique is
proposed which exploits the knowledge of both temporal and source
correlations. The JSCD technique is composed of two stages, which
are iteratively performed. First, a sum-product (SP) decoder is
serially concatenated with a BCJR decoder, where the knowledge of
source memory is utilized during {\em local (horizontal)
iterations}. Then, the estimate of correlation between the sources
is used to update the concatenated decoder during {\em global
(vertical) iterations}. Therefore, the correlation of the sources is
assumed as side information in the subsequent global iteration of
each concatenated decoder. From the simulation results of frame/bit
error rate (FER/BER), we note that significant gains are achieved
by the proposed decoding scheme with respect to the case where
the correlation knowledge is not completely utilized at the decoder.
\end{abstract}

\section{INTRODUCTION}
In {\em distributed source coding} problem, {\em Slepian-Wolf (SW)}
theorem states that using joint decoding achieves higher compression
rates compared to separate decoding of correlated sources
\cite{SW-73}. Numerous distributed compression methods have been developed for non-uniform sources or sources with memory
during recent years in the context of sensor networks, e.g., \cite{TRG-10a,DBRK-12}, and
references therein.

Although there are many efficient coding methods
for lossless compression, transmission of correlated sources over noisy
channels based on the SW cooperation has been investigated quite recently
\cite{BS-06,FZ-05,FZZ-07,DLM-06}.
To exploit the benefit of the SW cooperation over the relay
channels, {\em joint source-channel decoding (JSCD)} techniques have
been recently proposed based on turbo coding with significant bit
error rate (BER) performance gain \cite{AM-12, ZCAM-12}. Turbo-like
codes which are used for encoding of correlated sources are assumed,
having long block-lengths and hence the decoding latency is a
problem. Thus, these codes are not suitable in certain practical
applications due to decoding latency and/or power consumption
limitations.

Furthermore, most of the research results on transmission of
correlated bit sequences over noisy channels focus on sources
without {\em temporal correlation (source memory)}, and hence the
sources are assumed to be independent and identically distributed
(i.i.d) in the time domain. In \cite{FZ-05} and \cite{FZZ-07}, turbo
and low-density generator-matrix (LDGM) codes were proposed for
transmission of correlated binary i.i.d sources over additive white
Gaussian noise (AWGN) channel, respectively. Also in \cite{DLM-06},
a joint iterative decoder of LDPC codes was introduced for correlated
sources without memory. The decoding technique proposed in
\cite{DLM-06} employs a two-stage iterative decoder, between which
{\em local} and {\em global} iterations are performed.
In the local iterations, decoding is performed using the sum-product (SP)
algorithm \cite{MK-99}, while in the global iterations the estimate of
correlation between the sources is passed on to the SP decoder to
improve the decoding performance in the subsequent round of local
iterations.

The main objective of this paper is to propose a novel technique for
the transmission of two correlated binary sequences following
a first order Markov process, in contrast to \cite{FZ-05,FZZ-07,DLM-06} where the sources are i.i.d.
Relatively short-length (length < $10000$) LDPC codes
with systematic parity-check matrices are used for
the encoding of bit sequences to be transmitted because efficient
performance improvement can be expected with the help of the correlation
within and across the sequences in both the waterfall and error floor regions.

At the decoder side, the joint decoder proposed in this paper
exploits the information related to the temporal and source
correlations in two stages. A standard BCJR decoder \cite{BCJR-74} is
serially concatenated to a SP decoder for the efficient
utilization of the {\em temporal memory} structure inherent within
the output sequences of the source. Additionally, to exploit the
source correlation, the method proposed in \cite{DLM-06} is used,
where the {\em log-likelihood ratio (LLR)} of the local iterations
is updated according to the knowledge about the source correlation
estimated in the previous global iteration. Numerical results
demonstrate that significant BER performance gains are achieved by the proposed
JSCD method for the correlated binary sequences generated
by a Markov source.

The organization of the paper is as follows. In Section
\ref{sec::notation}, basic concepts and notations related to the
proposed algorithm are introduced. In Section \ref{sec::model}, the
proposed joint decoder structure is described in details. In Section
\ref{sec::simul}, simulation results and comparisons with
decoders that do not exploit the temporal and source correlations
are provided. Finally, Section \ref{sec::concul} concludes the
paper.
\section{PRELIMINARIES}
\label{sec::notation}
\subsection{Source Model}
We consider two binary sources, one of them is modeled as a binary Markov source whose sequences
are denoted by ${\bf b}^{1} =\{b_{1}^{1},b_{2}^{1},\ldots\}$, and the
sequences of another source, denoted by ${\bf
b}^{2}=\{b_{1}^{2},b_{2}^{2},\ldots\}$, is assumed to be erroneous
version of ${\bf b}^{1}$, corrupted by a binary noise with
occurrence probability $p$, where $b_{i}^{q}\in \{0,1\}$,
$q=1,2$. Therefore, $b_{i}^{2}=b_{i}^{1}\oplus z_i$, where $Z_i$ and
$\oplus$ denote a binary random variable with probability
$p_r(z_i=1)=p$ and a modulo-$2$ addition, respectively. The probability $1-p$
determines correlation ratio between the bit sequences of the two sources, i.e., the smaller the $p$ is,
the more correlated the sources are.

The binary Markov source considered in this paper is a stationary, ergodic,
state emitting Markov source $\{S_t\}$,
whose transition probabilities are described by the transition matrix
\begin{displaymath}
\Pi=[\pi_{ij}]=\left[\begin{array}{c c}
1-\alpha & \alpha \\
\beta & 1-\beta
\end{array}\right],
\end{displaymath}
where the transition probability is defined
\begin{displaymath}
\pi_{ij}\triangleq p_r\{S_t=j|S_{t-1}=i\},\quad i,\,j\in \{0,1\}.
\end{displaymath}

The stationary distribution of the source states is denoted by
$(\mu_0,\mu_1)$ where $\mu_0\triangleq p_r\{S_t=0\}\triangleq
1-\mu_1$. With the stationarity assumption, it can be easily shown
that $\mu_0$ and $\mu_1$ are
$\mu_0=1-\mu_1=\frac{\beta}{\alpha+\beta}$.
If $\alpha\ne \beta$, the source described above is an {\em asymmetric} binary Markov
source. 
Otherwise, when $\alpha=\beta\neq 1/2$ the source is a {\em symmetric} Markov source with a uniform
stationary distribution, i.e., $\mu_0=\mu_1=1/2$.

The {\em entropy rate} of stationary binary Markov source is
given by
\begin{equation}
H(s)=\mu_0h(\alpha)+\mu_1h(\beta), 
\end{equation}
where $h(p)=-p\log_2(p)-(1-p)\log_2(1-p)$ is the binary entropy
function \cite{CT-06}.

We will assume in this paper that the Markov source characteristics, i.e., $\alpha$ and $\beta$,
are known to the proposed decoder. Estimation of
Markov source parameters was discussed for unknown temporal correlation parameters in \cite{TRG-10b}.
\subsection{LDPC Codes}
Consider a binary LDPC code ${\cal C}$ represented by a Tanner graph
$G = (V_b \cup V_c , E)$, where $V_b = \{v_1, \ldots, v_n\}$ and
$V_c = \{c_1, \ldots, c_m\}$ are the sets of variable nodes and
check nodes, respectively, and $E$ is the set of edges.
Corresponding to $G$, we have an $m \times n$ parity-check matrix
$H=[h_{ij}]$ of ${\cal C}$, where $h_{ij} = 1$ if and only if (iff)
the node $c_i \in V_c$ is connected to the node $v_j \in V_b$ in
$G$; or equivalently, iff $\{v_j,c_i\} \in E$.
The code rate of ${\cal C}$ is
$R_c=1-\frac{\mathrm{rank}(H)}{n}$,
where $\mathrm{rank}(.)$ is the rank of its argument matrix in
$GF(2)$.

The {\em degree} of a variable or a check node is the number of nodes connected to it. If the nodes in the set $V_b$
and/or the nodes in the set $V_c$ have different degrees, the corresponding LDPC code is called {\em irregular}.
For an irregular LDPC code, the {\em degree distribution} of
variable nodes
is described by the polynomial, $\lambda(x)=\sum_{i=2}^{D_v} \lambda_i x^{i-1}$, where
$D_v$ is the maximum variable node degree
and $\lambda_i$ is
the fraction of the edges connected to the variable nodes of degree $i$.

\section{PROPOSED SYSTEM MODEL}
\label{sec::model}
\subsection{LDPC Encoder}
In our proposed scheme, two individual LDPC codes are used for
encoding of the correlated bits over independent AWGN channels.
The encoded bits are transmitted over the channels by using
binary phase-shift keying (BPSK) modulation.

To have an efficient BER performance, two types of optimized degree
distributions over AWGN channels with rates $0.5$ and $0.32$ are
used \cite{RSU-01}. The corresponding variable node degree
distribution for code rate $0.5$ is given by \cite[Table I]{RSU-01}
\begin{equation}
\lambda_A(x)=0.25105x+0.30938x^2+0.00104x^3+0.43853x^9,
\end{equation}
and for code rate $0.32$ is calculated as following
\begin{equation}
\lambda_B(x)=0.3127x+0.3582x^2+0.04x^6+0.2891x^9,
\end{equation}
both of which have maximum variable degree $10$.
A modified {\em progressive edge
growth (PEG)} method is used to construct relatively short-length parity-check
matrices with very low error-floor performance\cite{KAB-12}.

Just for the simplicity of the iterative decoders, we assume the
systematic parity-check matrices with equal rates for the both
channels.

\subsection{Theoretical Bound}
Let $H(s_1,s_2)$ denote the joint entropy of the first and the second sources.
According to the SW bounds, the total required rate to
transmit the correlated information sequences over the both channels
is given by
\begin{equation}
R=H(s_1,s_2)/(1/R_{c_1}+1/R_{c_2}),
\end{equation}
where $R_{c_i}$ for $i\in\{1,2\}$ is the code rate of the $i$-th channel, where the
terminologies and/or symbol notations follow \cite{FZ-05} and
\cite{FZZ-07}. The energy per source bit, $E_{so}$, can then be
related to the energy per information bit, $E_b$, and the energy
$E_s$ per sent symbol to be sent over the channels is given by
\begin{equation}
2E_{so}=H(s_1,s_2)E_b=(1/R_{c_1}+1/R_{c_2})E_s,
\end{equation}
where $E_s=1$ in BPSK modulation.

In general, the parity-check matrices of each channel can be different
in code rates. However, we only focus on the
cases that both channels use the same rate of the codes ({\em symmetric
case}), i.e., $R_{c_1}=R_{c_2}$, hence the {\em Shannon/SW} limit for
calculation of $E_{so}/N_0$ in the symmetric case is given by

\begin{equation}
(\frac{E_{so}}{N_0})|_{lim}=\frac{2^{H(s_1,s_2)R_{c_1}}-1}{2R_{c_1}},
\end{equation}
where $\sigma^2_n$ and $N_0=2\sigma^2_n$ respectively denote the noise
variance of the two AWGN channels, which is also
assumed to be equal, and the noise power spectral density.

Moreover, $H(s_1,s_2)=H(s_1)+h(p)$ where $H(s_1)$ is the entropy rate of the binary Markov source
and $h(p)$ is the entropy of the correlation parameter.

\subsection{Joint Source-Channel Decoder}
In the class of JSCD, which the proposed technique also belongs to,
the processing of correlation between the sources is performed at the decoder side.
Hence, the design of the joint decoder that can best exploit the
source-channel code properties plays very important role.
In the proposed decoder, the temporal and source correlations are exploited
in two consecutive iterations, these iterations are referred to as
local (horizontal) and global (vertical) iterations.
During the horizontal iterations, the systematic variable nodes have
LLR inputs from the received signals and a BCJR decoder matched to
the trellis diagram of the Markov source. The LLRs are exchanged
between variable and check nodes of the LDPC code until either no
more relevant improvement in LLR is achieved or a maximum number of
local iterations is reached. Then, the correlation between two
bit sequences is estimated at the end of the horizontal iterations and the
LLR associated to the source correlation is passed on to the
systematic variable nodes of each channel via the vertical iteration.
For the initialization, the LLRs to be used in the vertical
iteration are first set zero for the both decoders.

\begin{figure}[h]
\centering
\includegraphics[width = 0.45\textwidth]{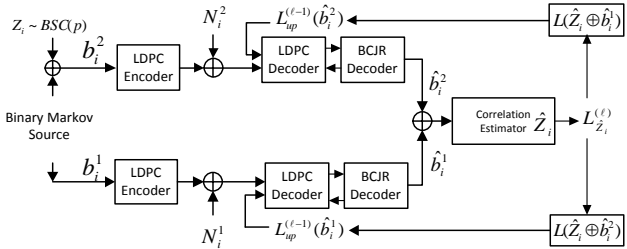}
\caption{Block diagram of the proposed JSCD technique for correlated
binary sequences generated by a Markov source.} \label{fig::blkdiag}
\end{figure}

In Fig. \ref{fig::blkdiag}, the block diagram of the proposed JSCD structure
is presented.

Let us consider the $j$-th local and the $\ell$-th global iteration of
the proposed decoder. Let $L_{v,c}^{(j)}$ and $L_{c,v}^{(j)}$
respectively denote the LLR messages passed from a variable node $v$
to a check node $c$ and from the $c$-th check node to the $v$-th variable
node at the $j$-th iteration. Also, $L_v^{cc}$ represents
the LLR obtained from the received signal to be input to the
variable node $v\in\{1,\ldots,n\}$.
Thus, $L_v^{cc}=2r_v/\sigma^2_n$
where $r_v$ is the received noisy signal. The index of the channels is omitted for
the simplicity, so far as no ambiguity is expected in the notations.

The process for LLR message updating in the variable nodes follows
the standard SP decoding with additional LLRs from the BCJR decoder
with a modification due to the source correlation, as detailed for
the expressions of Eq. \ref{eq::eq8}-\ref{eq::eq10}.
The information exchange between the SP and the Markov decoders does not contain the
information obtained by the previous iterations, and hence is {\em
extrinsic information}.

The LLR messages to be forwarded from the
$k$ systematic variable nodes to the corresponding check nodes are
given by
\begin{equation}
\label{eq::eq8}
L_{v,c}^{(j)}=L_v^{cc}+\sum_{c'\ne
c}L_{c',v}^{(j-1)}+L_{M,v}^{(j-1)}+L_{v,up}^{(\ell-1)},
\end{equation}
where $v\in\{1,\ldots,k\}$, $L_{M,v}^{(j)}$ and
$L_{v,up}^{(\ell)}$ denote the extrinsic LLR, sent from the
Markov decoder to the variable node $v$ and the LLR already updated according to the
correlation, respectively, in the $j$-th and the $\ell$-th iterations. It is
worth mentioning that summation is performed only on the check nodes
connected to the variable node $v$.

For parity variable nodes, the updated messages are obtained in the
same way as the standard SP decoder, and is represented by
\begin{equation}
\label{eq::eq9} L_{v,c}^{(j)}=L_v^{cc}+\sum_{c'\ne
c}L_{c',v}^{(j-1)},
\end{equation}
where $v\in\{k+1,\ldots,n\}$. The messages $L_{c,v}^{(j)}$ are
also updated in each local iteration exactly in the same way as the
standard SP decoder.

The input extrinsic LLR of the Markov decoder is calculated by
subtracting the extrinsic LLR of the Markov decoder, obtained as the
result of previous Markov decoding, from the {\em a posteriori} LLR,
obtained as the result of the current SP decoding, in each local
iteration. Thus, the sent LLR from a variable node $v$ to the Markov decoder in the
$j$-th iteration, denoted as
$L_{v,M}^{(j)}$, is calculated as
\begin{equation}
\label{eq::eq10}
L_{v,M}^{(j)}=L_v^{cc}+\sum_{c'}L_{c',v}^{(j-1)}+L_{v,up}^{(\ell-1)},
\end{equation}
where $v\in\{1,\ldots,k\}$.
Furthermore, the BCJR decoder which uses the trellis
diagram representing the Markov source is applied
to calculate the extrinsic
$L_{M,v}^{(j)}$ values by the
same equations represented in \cite{ZCAM-12}.

During the vertical iterations, the LLR $L(\hat{z}_v)$ of the binary
error sequence, $z_v$, is estimated. The
estimate of the binary error, $\hat{z}_v^{(\ell)}$, in the $\ell$-th
vertical iteration, is given by
\begin{equation}
\hat{z}_v^{(\ell)}= \hat{b}_v^{1,(\ell)}\oplus \hat{b}_v^{2,(\ell)},
\end{equation}
where $\hat{b}_v^{q,(\ell)}$, $q=1,2$, is the estimated source
(systematic) bits of the $q$-th channel obtained as the result of the
horizontal iterations. Hence, $\hat{b}_v^{q,(\ell)}$ is calculated as
\begin{equation}
\hat{b}_v^{q,(\ell)}= \left\{ \begin{array}{ll}
0 & \textrm{if $L_{v}^{q,(\ell)}\geq 0$}\\
1 & \textrm{if $L_{v}^{q,(\ell)} < 0$}\\
\end{array} \right. ,
\end{equation}
where
$L_{v}^{q,(\ell)}=L_v^{q,cc}+L_{M,v}^{q,(j_t-1)}+L_{v,up}^{q,(\ell-1)}+\sum_{c'}L_{c',v}^{q,(j_t-1)}$
and $j_t$ is the terminated horizontal iteration.
The LLR $L(\hat{z}_v^{(\ell)})$ of $\hat{z}_v^{(\ell)}$ is estimated
using the method presented in \cite{DLM-06}, as
\begin{equation}
L(\hat{z}_v^{(\ell)})=(1-2\hat{z}_v^{(\ell)})\frac{k-W_H}{W_H},
\end{equation}
where $k$ and $W_H$ denote the length of the original source packet and
Hamming weight of estimated binary error vector ${\bf
\hat{Z}^{(\ell)}}=\{\hat{z}_1^{(\ell)},\ldots,\hat{z}_k^{(\ell)}\}$.

Finally, according to the modulo-$2$ addition of the error sequence,
the updating LLRs of the first and the second concatenated
decoders are given by
\begin{align}
L_{v,up}^{1,(\ell)}&=2\,\mathrm{atanh}(\tanh(L(\hat{z}_v^{(\ell)})/2)\tanh(L_{v}^{2,(\ell)}/2)),\\
L_{v,up}^{2,(\ell)}&=2\,\mathrm{atanh}(\tanh(L(\hat{z}_v^{(\ell)})/2)\tanh(L_{v}^{1,(\ell)}/2)),
\end{align}
where $\tanh(.)$ and $\mathrm{atanh}(.)$ denote, respectively, the hyperbolic and inverse hyperbolic tangent functions.

\section{SIMULATION RESULTS}
\label{sec::simul}
In this section, we demonstrate the advantages
characteristics of the JSCD technique rather than using a single
structure which only exploits either the temporal or source correlations. The
distributed source-channel (DSC) coding was performed for symmetric
and asymmetric binary Markov sources with different code
rates over AWGN channels. In both examples, the maximum number of
local and global iterations for the proposed JSCD technique was set
at $50$ and $15$, respectively.

\begin{example}
\label{exm::ex1} In this example, a symmetric binary Markov source
with the state transition parameters $\alpha=\beta=0.1$ was
considered. For DSC coding of the correlated transmitted bits, two different
irregular LDPC codes with the degree distribution $\lambda_A$, the
rate $R_c=0.5$, and the block-length $n=4096$ were constructed by using
a modified PEG method \cite{KAB-12}.

The FER/BER performances of the transmitted source bits with
different decodings are presented in Fig. \ref{fig::ex1-1}, including
the proposed JSCD, the concatenated SP-BCJR decoding, the decoding
method proposed in \cite{DLM-06}, and the standard SP decoding.
It is found that roughly $1.3$ dB gain in terms of $E_{so}/N_0$ is
achieved by the proposed decoder with the correlation parameter $p=0.01$
compared to the case where only the temporal correlation of the source is used.
The gap between theoretical limits\footnote{The theoretical limits are obtained
using the same techniques described in \cite{FZ-05,ZA-06}.} and
our simulation results for the standard SP decoder, the decoder proposed
in \cite{DLM-06}, the concatenated SP-BCJR decoder and the proposed
decoder are $1.5$ dB, $3.5$ dB, $2.9$ dB and $4$ dB, respectively.

\begin{figure}[t]
\centering
\includegraphics[width = 0.45\textwidth]{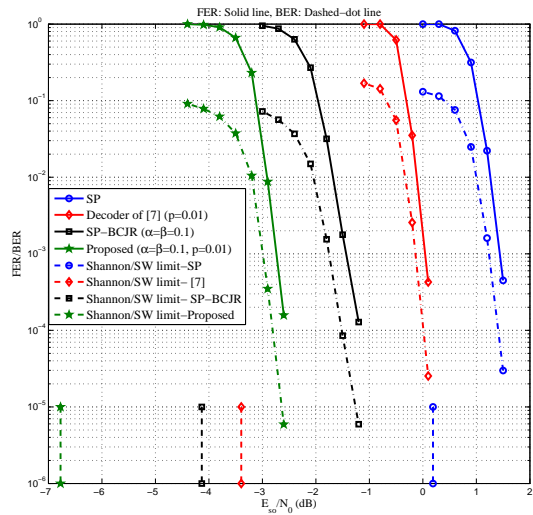}
\caption{FER, BER, Shannon/SW limit performances of various decoders for
correlated bit sequences with Markov source parameters $\alpha=\beta=0.1$, and
$p=0.01$
(Example \ref{exm::ex1}).} \label{fig::ex1-1}
\end{figure}

\begin{figure}[t]
\centering
\includegraphics[width = 0.45\textwidth]{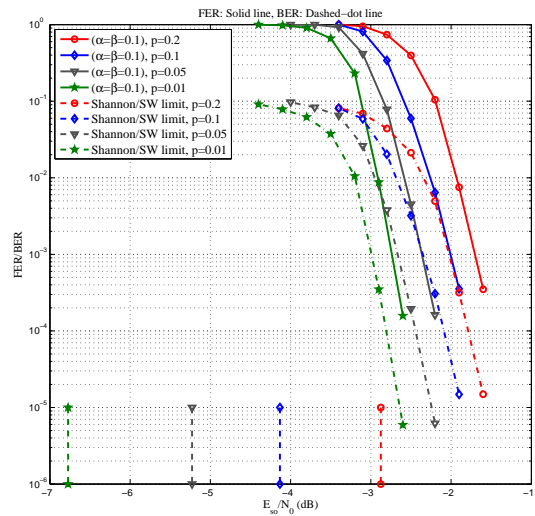}
\caption{FER, BER, and Shannon/SW limit performances of the proposed
decoder for different correlation values with Markov
source parameters $\alpha=\beta=0.1$ (Example \ref{exm::ex1}).}
\label{fig::ex1-2}
\end{figure}

The FER/BER curves of the JSCD method with different correlation
parameters and their respective Shannon/SW limits are shown in Fig.
\ref{fig::ex1-2}. The gap between the theoretical limits and the simulation
results for the JSCD with different correlation parameters
$p\in\{0.01,0.05,0.1,0.2\}$ are $\{4,2.9,2.2,1.5\}$ dB,
respectively. It is found that the more the correlation is between
the sources, the more the gap is between the theoretical limits and
the simulation results. The gap indicates that there will remain many
open questions for the optimal code design and decoding algorithm
development which are left as
a future study.
\end{example}

\begin{example}
\label{exm::ex2} The proposed JSCD technique was also performed for
an asymmetric binary Markov source with $\alpha=0.1$ and
$\beta=0.2$. Two different irregular LDPC codes were constructed by
the modified PEG method for encoding of correlated sources by using
the degree distribution $\lambda_B$ with the rate $R_c=0.32$, and the
block-length $n=6400$.

The FER/BER performance of the proposed decoder with the correlation parameter
$p=0.01$ is demonstrated in Fig. \ref{fig::ex2-1}. In this figure, the
FER/BER results of the SP decoder, the decoder proposed in
\cite{DLM-06}, and the concatenated  SP-BCJR decoder are also
depicted. As shown in Fig. \ref{fig::ex2-1}, more than $1.4$ dB
gain in terms of $E_{so}/N_0$ is achieved over the decoders that only use either the temporal
or source correlations. In this case, the gap between the theoretical limits and
the simulation results for the SP decoder, the decoder presented in
\cite{DLM-06}, the concatenated SP-BCJR decoder, and the proposed
decoder are
$1.5$ dB, $3.5$ dB, $2.7$ dB and $3.95$ dB, respectively.

Moreover, performances of the proposed decoder for different
correlation parameters, $p\in\{0.01,0.05,0.1,0.2\}$, and their
corresponding limitations are demonstrated in Fig. \ref{fig::ex2-2}.
The gap between the theoretical limits and the numerical results for
the correlation parameters described above are, respectively,
$\{3.95,2.9,2.2,1.5\}$ dB. It is obvious that the more the correlation is,
the more the gap is between the theoretical limits and the simulation
results just like the example \ref{exm::ex1}.

\begin{figure}[t]
\centering
\includegraphics[width = 0.45\textwidth]{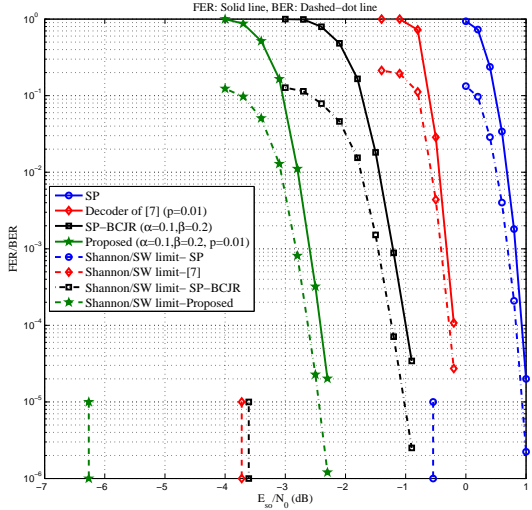}
\caption{FER, BER, Shannon/SW limit performances of various decoders for
correlated bit sequences with Markov source parameters $\alpha=0.1,\,\beta=0.2$, and
$p=0.01$ (Example \ref{exm::ex2}). } \label{fig::ex2-1}
\end{figure}

\begin{figure}[t]
\centering
\includegraphics[width = 0.45\textwidth]{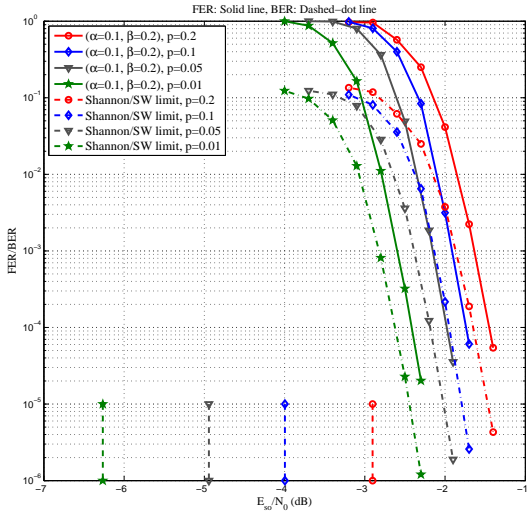}
\caption{FER, BER, and Shannon/SW limit performances of the proposed
decoder for different correlation values with Markov
source parameters $\alpha=0.1,\,\beta=0.2$ (Example \ref{exm::ex2}).}
\label{fig::ex2-2}
\end{figure}
\end{example}

\section{CONCLUSION}
\label{sec::concul} We have introduced an iterative joint
source-channel decoding technique for the transmission of two
correlated binary Markov sources. Each of the sources is independently encoded by
a relatively short-length irregular LDPC code. The
proposed decoder contains two concatenated SP-BCJR decoders, each
taking into account the temporal correlation (Markov memory
structure) over which extrinsic information is exchanged. Moreover,
the output of extrinsic LLRs from the Markov decoders is updated by
exploiting the source correlation during the global iterations. The
proposed JSCD technique outperforms the other decoders that do not
fully utilize the correlation information. Even though the proposed
decoding effectively utilizes temporal and source correlations,
there still remains a gap between the threshold $E_{so}/N_0$, obtained
by the simulations, and the theoretical limits. This indicates that there
are many open questions for the optimal code
design and decoding algorithm development which
are left as a future study.

\section*{Acknowledgment}
This work was supported by the Academy of Finland under SWOCNET project.

\end{document}